# SOA Embedded in BPM: A High Level View of Object Oriented Paradigm

Imran S.Bajwa

*Abstract*—The trends of design and development of information systems have undergone a variety of ongoing phases and stages. These variations have been evolved due to brisk changes in user requirements and business needs. To meet these requirements and needs, a flexible and agile business solution was required to come up with the latest business trends and styles. Another obstacle in agility of information systems was typically different treatment of same diseases of two patients: business processes and information services. After the emergence of information technology, the business processes and information systems have become counterparts. But these two business halves have been treated under totally different standards. There is need to streamline the boundaries of these both pillars that are equally sharing information system's burdens and liabilities. In last decade, the object orientation has evolved into one of the major solutions for modern business needs and now, SOA is the solution to shift business on ranks of electronic platform. BPM is another modern business solution that assists to regularize optimization of business processes. This paper discusses how object orientation can be conformed to incorporate or embed SOA in BPM for improved information systems.

## I. INTRODUCTION

TYPICALLY, web services are programs including methods that are exposed on web using XML at a particular URL. Moreover, these methods provide operational support by providing secure and reliable execution of the code. This operational support can help out in designing a BPM platform on various components: software, hardware, database, etc. The basic work flows across different business units are carried out by ERP. A modern business enterprise is battling at two principally different borders: Information Technology services and business processes. At one side information services are rigid and cohesive and at the other side business processes are not being dynamically optimized [2]. This vexing problem can be simplified by simply resolving the inundated issues by both sides as the borders of these two rivals are overlapped by each other. Following section clarifies the situation by plainly defining the issues.

This research work was conducted in the department of Computer Science & IT, The Islamia University of Bahawalpur and supported in part by the Higher Education Commission, Pakistan. This paper is the continuation of the series of publications under the research project of Natural Language Processing based Software Designing by the Information Processing Research Group (IPRG).
I. S. B. Author is Assistant Professor in the Department of Computer Science & IT, The Islamia University of Bahawalpur. He is regular member of IEEE, ACM, AAAI and IACSIT. (phone: +92 (062)9255466; e-mail: imran.sarwar@iub.edu.pk).

### A. Services based Business Enterprises

Following are some key features that led the business enterprises to find another solution in place of ERP.
- With the passage of time, where the number of units under a business enterprise have been increased at the same time the volume of these business units has also been increased radically.
- These days, business transactions are not as simple to make few queries to underlying database.
- With the increase in size and volume of business enterprises, the size and complexity of ERP has also boomed [3].
- On the other side, modern ERP solutions are overwhelming software packages from different venders to provide the maximum throughput to a modern business enterprise.
- From all these different packages, most of them are not nimble and responsive. Some have platform clashes and some others have conflicts with core tools and still some others have colloidal with the used information technologies [4].
- Still, there was another issue that business was conforming to online (web) grounds and ERP has not by-default support for this purpose.

So, the solution is to buy services from the venders not the software packages. SOA (Service Oriented Architecture) is composition of services with many more benefits. SOA has done a supported business and IT administrators by making the things simple and clear [5]. The reason of ultimate success of SOA has by-default potential to considerably modify the businesses associations and reform the whole industry.

### B. BPM for Business Enterprises

Typically, the major participants involved in a small business enterprise are vendors, IT infrastructure, vendors, database, customers, etc. Business Process Management (BPM) is the solution to synchronize communication among all these constituents [6]. By-definition, BPM is solution for dynamically optimizing the business processes. By using BPM in a business enterprise, manifold benefits can be ensued [7]:
- Dynamic changes in business processes can be tempt in the business enterprise.
- Incessant comprehension of the business processes in the business organization.
- Continuous monitoring of optimized processes is possible.
- All the business processes are mannered to be standardized.
- Multiple business processes can be integrated in complex but efficient manner.
- Space to resolve the tailbacks that ward off proficiency and competence.

The next section of this paper presents the object oriented characteristics of both BPM and SOA. In the later on sections, the relationship of BPM and SOA has defined and also the importance of BPM and SOA is justified. At the end, an object oriented based architecture to integrate SOA and BPM is also presented and some implementation constrains are also discussed in the last section.

## II. OO AND INFORMATION SYSTEMS

Object orientation has always been key part of software designing paradigm in last few decades. The features of object orientation; modular approach, reusability, generalization, adhesiveness in modules, etc have been premier reasons of success of object orientation [4]. Object orientation is not just concerned with the programming languages and coding but it provides a complete base to a business organization's concrete structure. A study has been presented in the next section to find out that how these object orientation features can help out to incorporate or embed SOA in BPM to get benefits of both [6].

### A. SOA and Object Orientation

SOA has enhanced ability to make dynamic changes by endorsing the modular approach, both in and across the software packages and services. Due to the modular approach, SOA not only facilitates the building of flexible information system applications but also integrates the complex and assorted IT technologies [8]. SOA is the high level view of the object orientation paradigm. An information system based on object orientation paradigm can be shifted on SOA with equal ease and competence. They are distinct features of OO that can support this counteractive shift. First, Object orientation inhales improved interpretability to induce harmony in business services. Object orientation supports better resource planning by increasing the associations among the IT and business resources [9]. By pursuing the modular approach, the organization remains impervious of dismals of the venders' services. OO supports increased and improved configuration of business and technology spheres. IT burden is reduced in terms of time, budget, workforce, etc. and agility among different organizational units is endorsed.

### B. BPM and Object Orientation

The modeling of business processes can also be carried out using object oriented analysis. Object orientation can help in modeling template processes and then using those templates to generate new processes in short time span [10]. The concept of reusability can also be augmented here to make the things simple and conspicuous. The procedure of modeling a business process on regular and dynamic basis becomes more refined and quick using objects oriented technology. Object orientation features in BPM can help in getting improved business agility. Agility in business processes can not simply come from automating them; it also requires a process-oriented approach to harmonize tasks between personnel of business and information technology units [11]. A business unit process can be managed, modeled, and improved with more simplicity and proficiency. In object oriented process modeling, modeling techniques are integrated with artifacts effectively supports each other. Using this approach, even very large and complex information systems can be quickly expressed and resolved. OO technology has been very successfully applied to several industry BPM projects [12].

## III. OO SUPPORT FOR SOA & BPM

In modern information system solutions, SOA and BPM both are considered critical constituents and both are used for their respective roles. Both disciplines have already established their significance and influence, working lonesome in the business organizations. These day larger organizations IBM [4], Microsoft [16], ORACLE [17], etc are thinking to emerge a combinatorial solution by using bests of both sides. Object orientation features are in both sides and can be binding factors of between BPM and SOA. Some integral possible binding features have been discussed in following:

- Objects based modular approach is the base of cohesive-less application components that us the key feature of SOA and also becomes the foundation of agility. Similarly BPM also steer implementation of the business processes as standalone components.
- Another trait enabled by object orientation is performance monitoring. In both SOA and BPM, it is expedient to monitor the performance on regular basis. This is possible due to object orientated nature as all business components are clearly defined and working separately just enabling counterparts to make the whole system a complete success.
- Object orientation helps SOA to support distributive nature and the feature of being distributive is common in both BPM and SOA. Typically, in a distributive environment, information is needed to contribute among various sites in a business enterprise with simplicity [13].
- With the help of objects, reusability is the main outgrowth of object orientation. By nature, both SOA and BPM support the reusability of modules and processes respectively. SOA supports loosely coupled modules that enable reuse in various applications [11]. Similarly, BPM as well chains the re-process of business rules and IT services by clearly defining boundaries of business services and logic. Feature of reusability not only lessens the complexity of the information system but also results in reduction of manufacturing, operational and management cost [14].
- Due to the support of modular approach in object orientation, another benefit is built-in support for change management. BPM and SOA by-nature manage and incorporate dynamic changes in business needs and requirements. BPM is convenient in doing so because of iterative nature and SOA as well has ability to adapt the alterations in the business logic.
- The last feature of SOA and BPM is iterative nature. Object Orientation once again enables both SOA and BPM to do so comprehensively. SOA use modular approach, hence it is easy for SOA to perform iterations. In the same way, BPM is also enables do to dynamically optimize the processes. [15] Another feature of continuous monitoring is also supported in models of both Service oriented Architecture and Business process Management.

## IV. EMBEDDING SOA IN BPM

The firs step to embed SOA in BPM is need of assimilation mechanism that supports loosely coupled applications otherwise the logic of a process will get hard-coded into a particular technology platform [18]. This is where standards-based service oriented architecture (SOA) comes in. An SOA provides the technical ability to create that process independence. SOA standards, such as Web Services, make information resources and task automation applications available yet loosely integrated for process designers to use and reuse [19]. Thus processes modeled with BPM tools can be rapidly implemented in production via SOA infrastructure. As shown in the figure 4, BPM is dynamic process for the automated process optimization and adaptation.

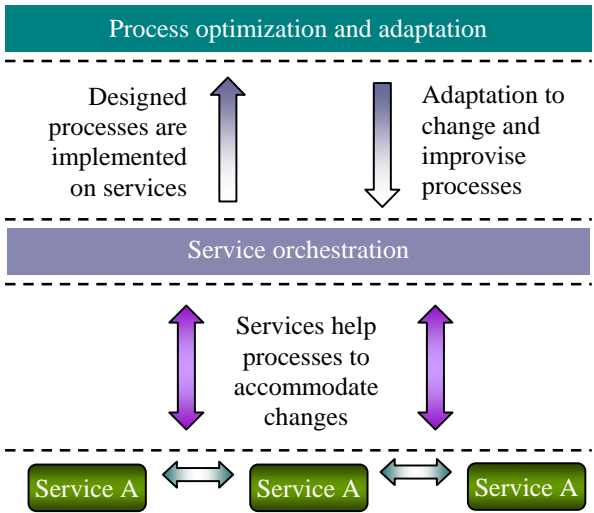

Fig. 1 BPM – Relationship of BPM and SOA

Service orchestration is main source of providing service agility. It provides an automated and settled arrangement for coordination of the service coordination. The services' management is also responsibility of service orchestration. All the above features help in supporting the reusability in components. SOA and BPM have adequate similarities that are required to build their composite architecture. For this purpose, the processes are implemented as services and in other words processes are mapped to the services [20]. A conventional form of BPM is shown in figure 2.

New and changed processes modeled in the BPM solution may be implemented in the enterprise infrastructure more rapidly because the SOA solution decouples the designed process from the specific implementation of particular applications that communicate only through a specific integration solution. The co-existence of SOA and BPM is need of modern business needs [21]. The joint venture of SOA and BPM is going to become a reality by converging SOA and BPM. Convergence of SOA and BPM is possible because the core functionality of the business processes in implemented through the IT services. So there is needed to embed the IT services somewhere in Business processes that are appropriate for both BPM and SOA.

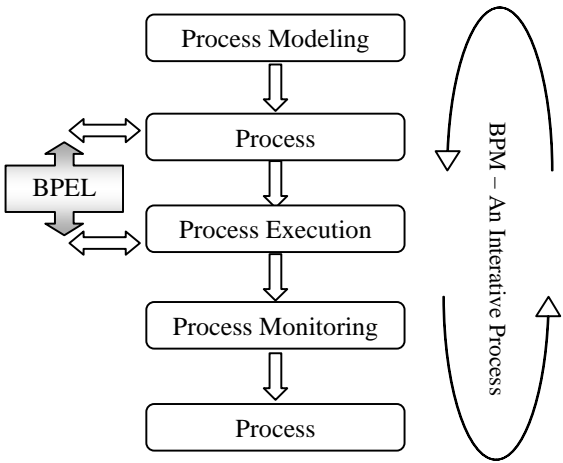

Fig. 2 BPM – Five Step Process Modeling

BPM starts from assessments of the requirements and modeling a new or existing process. After modeling, processes are implemented. Process execution is the next phase. BPEL (Business Process Execution Language) is the layer that helps in executing the business processes according to the business process logic [22]. Last step is to continuously monitor the business processes and optimize them if some alteration updation is required. Conventionally, BPM is a discipline that can handle issues: document processing, accommodating dynamic changes, monitoring processes' evaluating performance, etc [23].

On the other side, SOA has the built-in ability to considerably amend the businesses associations and reorganize information systems [13]. SOA allows to compose the business processes and services together; across application and organizational boundaries. Following is framework of BPM after embedding SOA.

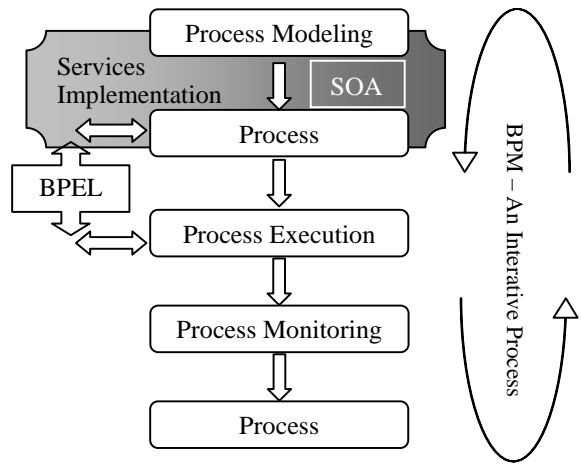

Fig. 3 BPM after embedding SOA

In the above five step BPM model, SOA is embedded in-between the phase of process modeling and process development phase. The responsibility of SOA is to accommodate the changes made in the modeled processes and then reflects those changes in developed processes. Implementation of the processes takes place by using underlying services. This integration of processes and IT services allows services to be used as reusable components that can be orchestrated [11] to support the needs of dynamic business processes.

V. PROPOSED FRAMEWORK

The previous section describes the basis that can be helpful in embedding SOA in BPM. The ultimate goal of this study was to emerge a single solution instead of two separate half-benefited solutions. Services are he main constituents of SOA. Need id to make services open for the business processes to make up the functionality of a business enterprise. Moreover, it should be insured that available services are also accessible to the employees, partners, and suppliers via the web [10]. In the presented framework, SOA is embedded in conventional model of BPM. The proposed framework's architecture is shown in figure 4.

- In the proposed framework, at the top, there is a user interface layer. This layer is used to receive user queries and hand over to the respective business processes. After completing the functionality the same layer also gives the reply back to the user.

- Next are the business processes. Different processes perform various functionalities. Functionality of each business process is implanted by using one or more underlying services. These business services provide new services interfaces based on enterprise semantic and functional requirements and also help to map them according to the existing system.

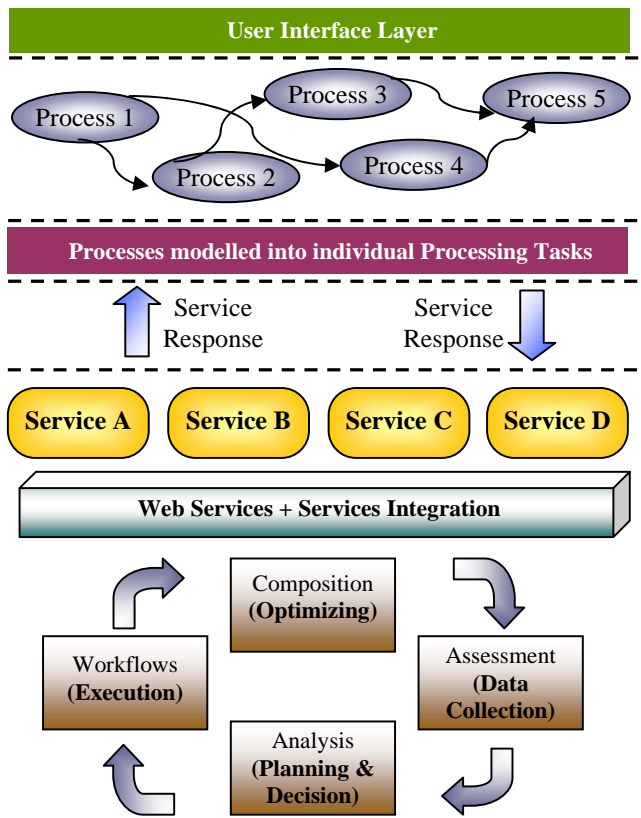

Fig. 4 Framework - SOA embedded in BPM

- After modeling the processes are discrete into individual processing tasks. These processing tasks are implemented using the IT service.
- Next in the framework are actual services. These services are independent of the underlying platform that consists of database, the programming languages and operating system. As, SOA is the combination of technologies that are required to enable the migration of inflexible IT functions into merged, loosely coupled and on-demand services. Following is the proposed combined architecture of BPM and SOA.
- Below services layer, service assembly and service integration has been provided for the sake of service agility. This layer provides distinct features to business services like automated arrangement, coordination and management are provided among the complex automated systems, middle-ware and services. These features also result in reusable software components that also support agility in the underlying information system. The typical support of web services is also available at the same layer.

At the bottom there is the complete life cycle of service creation is provided. It is a four step procedure including collection of data for assessment of the required service. After assessment, the plans are finalized and finally the workflows are implemented to execute the services. The implemented services are commonly monitored for optimization.

## VI. LITERATURE REVIEW

The use of BPM in concert with SOA – a perfectly aligned partnership of Business and IT investments – is the fast path to ensuring true business agility. BPM provides a wonderful abstraction for building business systems [16]. But all too often BPM is used to build higher level, more efficient, but nonetheless silo applications rather than contributing to an overall flexible, agile enterprise. This is where SOA comes in. SOA provides the application platform to bridges to the business processes and the operational resources [12]. Together, BPM and SOA provide a perfect combination for enterprise computing. BPM provides the higher-level abstraction for defining businesses processes, as well as other important capabilities of monitoring and managing those processes [9].

The BPM-SOA partnership allows services to be used as reusable components that can be orchestrated to support the needs of dynamic business processes [3]. The combination enables businesses to iteratively design and optimize business processes that are based on services that can be changed quickly, instead of being 'hard-wired'. A critical success factor of SOA-BPM is the adoption of industry recognized technology standards [11], which allow the architecture to be portable and executable in almost any chosen hardware and software environment, eliminating the need to be tied to any specific vendor. Colleen Frye [4] says that "BPM is a small fish inside the belly of the SOA whale…" In the same article, Colleen says also that "BPM and SOA are two sides of the same coin; joined at the hip". Mike Rosen [9] thinks that "BPM and SOA provide a perfect combination for enterprise computing". Ismael Ghalimi [6] says that "BPM is SOA's killer application and SOA is BPM's enabling infrastructure." According to Ismael, BPM cannot work together but there are many similarities in both SOA and BPM. BPM and SOA in combination help and facilitate the next phase of business process evolution – going from merely automating repeatable processes to flexible automation of dynamic processes [17].

ACKNOWLEDGMENTS

This research is part of the research work funded by HEC, Pakistan and was conducted in the department of Computer Science & IT, IUB, Pakistan. This article was really difficult to produce as the domains of BPM and SOA are still not properly defined and their integration is still in infant stage. I had to take many ideas from the authors like Isamael Ghalimi [6], Michael Madsen [11], Jesper Joergensen [13], Mike Rosen [9], etc.

**Author's Biography**

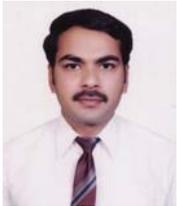

Imran Sarwar Bajwa is Assistant Professor in the Islamia University of Bahawalpur. He has been teaching and doing research in various universities in Pakistan for last 6 years. He has also been part of faculty and active member of research community in the University of Coimbra, Portugal in 2006-07. His major areas of research are Natural Language Processing, Information Retrieval (text & image) and Information Systems. He is member of different professional societies i.e. IEEE, AAAI, IASCIT, ACA, EATCS, SCTA, etc. He is also active reviewer of many journals. Following are his some major publications:

[1] "Web Layout Mining (WLM): A Paradigm for Intelligent Web Layout Design", *Egyptian Computer Science Journal*, Vol. 29, No. 2, May 2007

[2] "Database Interfacing using Natural Language Processing", European Journal of Scientific Research, Vol. 20 No.4 (2008), pp.844-851

[3] "Feature Based Image Classification by using Principal Component Analysis", ICGST - Journal of Graphics Vision, and Image Processing, Vol. 09 No.2 (2009) pp 11-17